\documentstyle[eqsecnum,preprint,aps,prd,epsfig]{revtex}
\preprint{DAMTP R-97/18, UCSBTH-97-10}
\newenvironment{abs}{\begin{center}\begin{minipage}
{0.9\textwidth}}{\end{minipage}
\end{center}}
\date{\today}
\draft
\tighten
\begin{document}
\def\sqr#1#2{{\vcenter{\hrule height.3pt
      \hbox{\vrule width.3pt height#2pt  \kern#1pt
         \vrule width.3pt}  \hrule height.3pt}}}
\def\square{\mathchoice{\sqr67\,}{\sqr67\,}\sqr{3}{3.5}\sqr{3}{3.5}}
\def\today{\ifcase\month\or
  January\or February\or March\or April\or May\or June\or July\or
  August\or September\or October\or November\or December\fi
  \space\number\day, \number\year}

\def\Bbb{\bf}

%%%%%%%%%%%%%%%%%%%%%%%%%%%%%%%%%%%%%%%%%%%%%%%%%%%%%%%%%%%%%%%%%
%%%% Beware of the kludge to put indices on the author names %%%%
%%%%%%%%%%%%%%%%%%%%%%%%%%%%%%%%%%%%%%%%%%%%%%%%%%%%%%%%%%%%%%%%%

\title{Can extreme black holes have (long) Abelian Higgs hair?}

\author{A. Chamblin{$^{1,4}$}, J.M.A. Ashbourn-Chamblin{$^{2}$},
R. Emparan{$^3$}, A. Sornborger{$^4$}}

\address {\qquad \\ {$^1$}Institute for Theoretical Physics,\\
University of California,\\
Santa Barbara, California 93106-4030, U.S.A.
\qquad \\ {$^2$}Wolfson College,\\
University of Oxford,\\
Oxford OX2 6UD, England
\qquad\\{$^3$}Dept. of Physics,\\
University of California,\\
Santa Barbara, CA 93106, U.S.A.
\qquad\\{$^4$}DAMTP,\\
Silver Street,\\
Cambridge, CB3 9EW, England
}
\maketitle

\vspace*{0.3cm}

\begin{center}
{Abstract}\\
\end{center}

\begin{abs}
{\small It has been argued that a black hole horizon can support 
the long range fields of a Nielsen-Olesen string, and that one 
can think of such a vortex as black hole ``hair''. In this paper, 
we examine the properties of an Abelian Higgs vortex in the presence 
of a charged black hole as we allow the hole to approach extremality.  
Using both analytical and numerical techniques, we show that 
the magnetic field lines (as well as the scalar field) of the vortex
are completely expelled from the black hole in the extreme limit.  
This was to be expected, since extreme black holes in 
Einstein-Maxwell theory are known to exhibit such a ``Meissner 
effect'' in general. This would seem to imply that a vortex 
does not want to be attached to an extreme black hole.  We calculate the 
total energy of the vortex fields in the presence of an extreme black hole.
When the hole is small relative to the size of the vortex, it is 
energetically favoured for the hole to remain inside the vortex region,
contrary to the intuition that the hole should be expelled.
However, as we allow the extreme horizon 
radius to become very large compared to the radius of the vortex, 
we do find evidence of an instability.  This proves
that it is energetically unfavourable for a thin vortex 
to interact with a large extreme black hole.  This would seem to 
dispel the notion that a black hole can support `long' abelian Higgs 
hair in the extreme limit. We show that these 
considerations do not go through in the near extreme limit. Finally, we
discuss whether this has implications for strings that end at black holes.}\\
\end{abs}


\section{Introduction}

Black hole `hair' is defined to be any field(s) associated with a stationary
black hole configuration which can be detected by asymptotic observers
but which cannot be identified with the electromagnetic
or gravitational degrees of freedom.  Back in the heyday of black hole
physics a number of results were proven (\cite{israel}, \cite{wald1}, 
\cite{carter})
which seemed to imply that black holes `have no hair'.  
Put more colloquially, these results implied that given certain
assumptions the only information
about a black hole which an observer far from the hole can determine
experimentally is summarized by the electric charge, magnetic charge, angular
momentum and mass of the hole.  
Such uniqueness results are referred to 
as `no hair' theorems.  These celebrated results would seem to imply that
a black hole horizon can support only these limited gauge charges; for a long
time people thought that other matter fields simply could not be associated
with a black hole.  However, this prejudice was to some extent discredited
when various authors \cite{group}, using numerical techniques,
discovered black hole solutions of the Einstein- Yang-Mills equations 
that support Yang-Mills fields which can be detected 
by asymptotic observers (these papers extended the earlier work of
Bartnik and McKinnon, who found globally regular finite energy solutions
in EYM theory without horizons \cite{bart}); 
one therefore says that these black holes are 
{\it coloured}.
Once the solutions of \cite{bart} were discovered, it wasn't long before
other people were finding similar solutions in Einstein-non-Abelian
gauge systems \cite{smol}.

However, these exotic solutions do not impugn the original no hair results
since all such solutions are known to be unstable (see e.g. \cite{biz}).
Since the original no hair theorems assumed a stationary
picture they simply do not apply to coloured holes. On the other hand,
coloured holes do still exist and so they are said to `evade' the usual
no hair results.
These results teach us that we have to tread carefully when
we start talking about black hole hair. 

There are other amusing tricks which allow one to evade no hair theorems.
For example, the reader will recall \cite{met} that in string theory the 
Einstein equations are induced from the low energy effective field theory
only to `zeroth' order in ${\alpha}^{\prime}$, where ${\alpha}^{\prime}$
denotes the Regge slope of string theory.  If you include the order
${\alpha}^{\prime}$ corrections then you get curvature `squared' terms
in the Lagrangian (you also get the usual dilatonic terms).  It turns out 
\cite{kanti} that black hole solutions in such a curvature squared, 
higher derivative theory of gravity can support non-trivial dilatonic 
configurations outside the horizon, and so they are said to possess 
`dilatonic' hair.  Again, these results do not actually contradict the 
original no hair theorems since they only apply in exotic situations.

What these results teach us is that we have to tread very carefully whenever
we start talking about black hole hair.  We will stick with our definition of 
hair as any property which can be measured by asymptotic observers. 
Furthermore, we shall follow \cite{ana} and use the term `dressing' for the
question of whether or not fields actually live on the horizon.

With all of this in mind, we now want to analyze the extent to which hair
is present in situations where we allow the topology of some field configurations
to be non-trivial; in particular, an interesting question is whether or not
topological defects, such as domain walls, strings, or textures 
\cite{review}, can act as `hair' for a black hole. 
In \cite{ana} evidence was presented that a 
Nielsen-Olesen ($U(1)$) vortex can act as `long' hair for a Schwarzschild
black hole.  More precisely, in \cite{ana} the authors studied the problem
of whether or not such a vortex can exist on a Schwarzschild black hole
background (neglecting at first the gravitational back reaction); 
they presented analytical and numerical evidence for such a solution.  
They went on to include the gravitational back reaction of a single 
thin vortex and managed to rederive the `AFV' metric (\cite{aryal}), 
which is a solution meant to model a cosmic string threading through 
a Schwarzschild black hole (i.e., the AFV solution is just a conical 
defect centered on a black hole). Thus, they were able to argue that 
the AFV solution truly is the `thin vortex' limit of a `physical' 
vortex-black hole configuration.  Using all of these results, they 
concluded with an argument that the abelian Higgs vortex is
{\it not} just dressing for the Schwarzschild black hole, but rather that 
the vortex is truly hair, that is, a property of the black hole which can
be detected by asymptotic observers.

In this paper, we extend the analysis of \cite{ana} and allow the black hole
to be charged.  That is to say, we consider the problem of an abelian Higgs
vortex in the Reissner-Nordstrom background.  In order to `turn up' the 
electric charge of the hole, we have to allow for the presence of two
$U(1)$s (one $U(1)$ is where the charge of the hole lives and the other
$U(1)$ is the symmetry spontaneously broken in the ground state); otherwise,
the charge would be screened.  We find that the results of \cite{ana}
are reproduced when the charge of the hole is very small relative to the 
mass.  However, as we increase the charge and the hole approaches extremality
we find that something very remarkable happens.  In the extreme limit, {\it all}
of the fields associated with the vortex (both the magnetic and scalar degrees
of freedom) are expelled from the horizon of the black hole.  We present
dramatic numerical evidence that the magnetic and scalar fields always 
`wrap around' the horizon in the extremal limit.  This behaviour was expected,
given that extreme black holes in Einstein-Maxwell-dilaton theories generically
display such a `Meissner effect', and so can be thought of as `superconductors'
(a deeper analysis of the superconducting properties of extremal black holes
and p-branes in Kaluza-Klein and string theories will be given in \cite{cham}).

We go on to calculate the total energy present in the electromagnetic field
(of the vortex) as we allow the extreme black hole to become very large
compared to the size of the vortex, and we find an instability.  Put more 
simply, for black holes large compared to the vortex radius
the energy of a vortex which does {\it not} wrap the hole (i.e., with the 
black hole outside the vortex) is
{\it much} less than the energy of a vortex which does wrap the hole.  It is
therefore energetically unfavourable for the vortex to interact with the hole,
and indeed the vortex will want to `slide' off of the hole. Thus, in the 
thin vortex limit a vortex does not want to be attached to an extreme black 
hole. It follows that the vortex cannot in any way be thought of as a 
`property of the black hole which can be measured at infinity'; 
in other words, an abelian Higgs vortex is not hair for an extreme black hole. 
Curiously, the expulsion of the vortex does not proceed gradually as the
black hole approaches extremality; rather, we have found numerical
evidence that a non-extreme black hole is always pierced by a vortex, 
no matter how close to extremality it is. We conclude with a discussion of
whether our results have 
implications to scenarios involving strings ending on black 
holes, in particular, the snapping of strings by the formation of 
black hole pairs.

\section{The Nielsen-Olesen vortex in the presence of a charged black hole}

In this section we analyze the Nielsen-Olesen equations for an 
abelian Higgs vortex \cite{no}
in the Reissner-Nordstrom background. Since we want to provide some 
continuity with 
the study of Ach\'ucarro et al.\ \cite{ana}, which in some respects we 
generalize, we will present our analysis in a form and notation that
closely parallel theirs.

Our treatment of the black hole/string vortex system involves a clear
separation between the degrees of freedom of each of these objects. 
The action takes the form
\begin{equation}
S = S_1  + S_2,
\end{equation}
where the first term is an Einstein-Hilbert-Maxwell action,
\begin{equation}\label{ehm}
S_1 = {1\over 16\pi G}\int d^4 x \sqrt{-g} \left( -R - 
{1\over 4}{\cal F}^2\right),
\end{equation}
and the second describes an abelian Higgs system minimally coupled
to gravity, 
\begin{equation}\label{abhiggs}
S_2 = \int d^4 x \sqrt{-g} \left( D_\mu\Phi^\dagger D^\mu\Phi -
{1\over 4 e^2} F^2 - {\lambda\over 4}(\Phi^\dagger\Phi - \eta^2)^2\right).
\end{equation}
The matter content of the abelian Higgs system consists of 
the complex Higgs 
field, $\Phi$, and a $U(1)$ gauge field with strength, $F_{\mu\nu}$, 
and potential $A_\mu$. Both the Higgs scalar and the gauge field 
become massive in the broken symmetry phase. 
They are coupled through the gauge covariant derivative
$D_\mu=\nabla_\mu + iA_\mu$, where $\nabla_\mu$ is
the spacetime covariant 
derivative. As in \cite{ana}, we choose metric signature $(+---)$. 

The degrees of freedom in $S_2$ will be treated as `test fields', i.e.,
their energy momentum tensor is supposed to yield a negligible contribution 
to the source of the 
gravitational field. The latter, instead, affects the propagation of
the fields $\Phi$ and $A_\mu$: an exact solution of the 
Einstein-Maxwell equations from $S_1$ will be plugged into the Abelian 
Higgs action $S_2$ as a fixed, background metric $g_{\mu\nu}$. 
Notice that we have two different gauge fields, ${\cal F}$
and $F$, and each is treated in a very different manner.
It is only $F$ that couples to the Higgs field and is therefore subject 
to spontaneous
symmetry breaking. The other gauge field, ${\cal F}$, could be thought of as 
the free, massless Maxwell field of everyday experience; apart from modifying
the background geometry, its dynamics
will be of little concern to us here.
Notice that whereas we treat $F$ as a test field, the backreaction of 
${\cal F}$ on the geometry will be fully accounted for.

The parameter
$\eta$ is the energy scale of symmetry breaking and $\lambda$ is the Higgs
coupling. These can be related to the Higgs mass by  $m_{\rm Higgs} = 
\eta\sqrt{\lambda}$. There is another relevant mass scale, i.e.,
that of the vector
field in the broken phase, $m_{\rm vector}= \sqrt{2} e\eta$. 
On length scales
smaller than $m_{\rm vector}^{-1}$, $m_{\rm Higgs}^{-1}$, the vector and 
Higgs fields behave as essentially massless.
It is also convenient to define the Bogomolnyi parameter
$\beta =\lambda/2e^2= m^2_{\rm Higgs}/m^2_{\rm vector}$. 

The action (\ref{abhiggs}) has a $U(1)$ invariance realized by
\begin{equation}
\Phi\rightarrow \Phi e^{i\Lambda (x)},\qquad A_\mu\rightarrow 
A_\mu - \nabla_\mu\Lambda(x),
\end{equation}
which is spontaneously broken in the ground state, 
$\Phi = \eta e^{i\Lambda_0}$.
Besides this ground state, another solution, the vortex, is present when 
the phase of $\Phi (x)$ is a non-single valued quantity. To better describe
this, define the real fields $X$, $P_\mu$, $\chi$, by
\begin{equation}
\Phi = \eta X e^{i\chi},\qquad 
A_\mu = P_\mu -\nabla_\mu\chi.
\end{equation}
A vortex is present when
$\oint d\chi = 2\pi N$, the integer $N$ being called 
the winding number of the vortex. If $N\neq 0$, and if
the spatial topology is trivial,
then, by continuity, the integration loop 
must encircle
a point of unbroken symmetry ($X = 0$), namely, the vortex core.

The Euler-Lagrange equations that follow by varying $X$ in the action
(\ref{abhiggs}) are
\begin{equation}\label{noeq1}
\nabla^2 X - X P_\mu P^\mu + {\lambda\eta^2\over 2}X(X^2-1) =0,
\end{equation}
while by varying $A_\mu$ one finds
\begin{equation}\label{noeq2}
\nabla_\mu F^{\mu\nu} + 2e^2\eta^2 X^2 P^\nu =0.
\end{equation}
The field $\chi$ is not dynamical. In flat space, vortices of 
Nielsen-Olesen type \cite{no} appear as cylindrically 
symmetric solutions,
\begin{equation}
\Phi = X(r_c) e^{iN\varphi},\qquad P_\varphi = N P(r_c),
\end{equation}
$r_c$ being the cylinder radial coordinate, and all other components 
of $P_\mu$ being zero.
We will be concerned, however, not with flat space, but
with another solution of the Einstein-Maxwell theory (\ref{ehm}), namely,
the Reissner-Nordstrom black hole
\begin{eqnarray}
ds^2 &=& V dt^2 - {d\rho^2\over V} - \rho^2(d\theta^2 + 
\sin^2\theta d\varphi^2),\\
V &=& 1 - {2 G m\over \rho} + {q^2\over \rho^2},\nonumber
\end{eqnarray}
(the charge $q$ is measured here in geometrical units), which
is not cylindrically symmetric. This makes the analysis of the 
solutions somewhat more complicated. 

It will be convenient to rescale the
radial coordinate and black hole
parameters by the Higgs wavelength to work with the non-dimensional variables 
$(r,\;E,\;Q)=\eta\sqrt{\lambda}(\rho,\; G m,\;q)$.
In terms of these variables,
\begin{equation}
V= 1 - {2 E\over r} + {Q^2\over r^2}.
\end{equation}
We stress that the charge $Q$ of the black hole, which couples to the field
$\cal F$, 
is unrelated to the abelian gauge field $F$ associated with the vortex. 
$Q$ can be primarily thought of as a parameter that allows us to 
modify the background geometry, in particular, to consider 
the extremal black hole backgrounds described below.

The Reissner-Nordstrom black hole has inner and 
outer horizons where $V(r)=0$. We will only be interested in the outer 
horizon, which is at radius
\begin{equation}
r_+ = E +\sqrt{E^2-Q^2}.
\end{equation}
The horizon exists as long as $E\geq |Q|$, otherwise one finds a naked 
singularity. 
If the inequality is saturated, $r_+ = E = |Q|$,  then $V(r)$ has
a double zero at $r_+$, 
and the black hole is said to be extremal.

Return now to the equations of the vortex. One can consistently take
\begin{equation}
X= X(r,\theta),\qquad P_\varphi = N P(r,\theta),
\end{equation} 
which simplifies the equations of motion (\ref{noeq1}), (\ref{noeq2}), 
to the form
\begin{equation}\label{rnoeq1}
-{1\over r^2}\partial_r( r^2 V\partial_r X) - {1\over r^2\sin\theta}
\partial_\theta(\sin\theta\partial_\theta X) + {1\over 2} X(X^2 -1)
+ {N^2 X P^2\over r^2\sin^2\theta} =0,
\end{equation}
\begin{equation}\label{rnoeq2}
\partial_r(  V\partial_r P) + {\sin\theta\over r^2}\partial_\theta\left(
{\partial_\theta P\over \sin\theta}\right) - {X^2 P\over \beta} =0.
\end{equation}
In this generic form these equations allow us to recover two interesting 
situations as limiting cases. First, when
$\beta\rightarrow \infty$  the Higgs field decouples. In this
situation we would be essentially studying a free Maxwell test field in 
the Reissner-Nordstrom
background. The complementary situation arises when $P=1$ (a constant)
throughout the space: this would be
a global string, i.e., without any local gauge dynamics, in the presence of
the charged hole. 

The equations (\ref{rnoeq1}), (\ref{rnoeq2}), are, in general,
rather intractable in exact form and we will need to resort to approximation 
methods. In the next section, we will solve the equations numerically
and study configurations with arbitrary relative sizes
of the black hole/vortex radii.
{}For the remainder of this section we will describe an analytical solution
of these equations for the case
where the black hole is small relative to the vortex size. 
In the units
we are using the radius of the flux tube is $r\sim \sqrt{2N}\beta^{1/4}$ 
for $N\gg 1$. 
Thus we will require
$\sqrt{N}\gg E$. This sort of large $N$ limit was first employed to 
obtain analytical
results in \cite{cinderella}. The results
we obtain in this way will be consistent with our numerical solutions in the
next section.

Well inside the core of the vortex the gauge symmetry remains essentially
unbroken. Thus, we expect $X\approx 0$, or better, $X^2/\beta\approx 0$. 
It is not difficult to see that, within the approximation considered,
one can consistently neglect the last term in the equation (\ref{rnoeq2})
and then attempt to solve
\begin{equation}\label{rnfree}
\partial_r(  V\partial_r P) + {\sin\theta\over r^2}\partial_\theta\left(
{\partial_\theta P\over \sin\theta}\right) \approx 0.
\end{equation}
{}For the Schwarzschild background ($Q=0$), a solution is provided by 
$P-1 \propto r^2\sin^2\theta$. This suggests that we
try the ansatz
\begin{equation}
P \approx 1 + a(r)\sin^2\theta.
\end{equation}
The equation we must solve now is
\begin{equation}
(r^2 -2 E r + Q^2) a'' + 2\left( E-{Q^2\over r}\right) a' - 2a =0,
\end{equation}
which admits the solution
\begin{equation}
a(r) = -p(r^2 -Q^2),
\end{equation}
hence
\begin{equation}\label{pfield}
P\approx 1- p(r^2 -Q^2)\sin^2\theta.
\end{equation}
Here $p$ is an integration constant equal to twice the magnetic field
strength at the center of the core. We have also chosen the parameters 
in order to have 
$P\rightarrow 1$ at the string axis $(\theta =0,\pi)$.
{}Far from the black hole, but still
inside the vortex, we can perform an analysis similar to that in 
\cite{cinderella} to show that
\begin{equation} 
p\approx {1\over 2N\sqrt{\beta}}.
\end{equation}
Large $N$ thus means small $p$.

Now we have to solve the equation for the Higgs field $X$, (\ref{rnoeq1}). 
{}Following \cite{cinderella},
we set $X=\xi^N$ and expand in powers of $1/N$.
This yields
\begin{equation}
V\left({\partial_r \xi \over \xi}\right)^2 + 
{1\over r^2}\left({\partial_\theta \xi \over \xi}\right)^2 = 
{P^2\over r^2\sin^2\theta}  +O(1/N^2).
\end{equation}
To be consistent we must neglect the terms proportional to $p^2$, since as
we have seen they would contribute to $O(1/N^2)$.\footnote{
This limits the 
validity of the solution to distances $r\sin\theta$ sufficiently smaller
than $\sqrt{2N}$.} 
Having done this, the equation becomes separable and can be solved in the form
$\xi = b(r)\sin\theta$, where $b$ must satisfy
\begin{equation}
{b'\over b} = {1 - p(r^2 -Q^2)\over r\sqrt{V}}.
 \end{equation}
This is integrated to yield
\begin{equation}\label{bsol}
b(r) = k (r - E +r\sqrt{V}) e^{-{p\over 2}(r^2+ 3 Er)\sqrt{V} - {3\over 2}
p (E^2 - Q^2)\ln (r-E + r\sqrt{V})}
\end{equation}
($k$ is another integration constant; its precise value is 
irrelevant for our purposes). From here we get $X$ as
\begin{equation}\label{xfield}
X\approx b^N(r)\sin^N\theta.
\end{equation}

Eqs.\ (\ref{pfield}), (\ref{bsol}) and (\ref{xfield}), constitute our solution 
describing a ``test vortex''
living in the background of a charged black hole that sits well within the
vortex core. The presence of charge induces a number of qualitative changes 
in the picture described in \cite{ana} (the neutral case).
To start with, notice that the distance at which $P\approx 0$, which roughly
defines the thickness of the vortex, is
\begin{equation}
r\sin\theta \approx\sqrt{{1\over p} + Q^2\sin^2\theta}\approx {1\over\sqrt{p}}
\left( 1
+ {p\over 2} Q^2\sin^2\theta\right),
\end{equation}
so we see that, compared to the neutral black hole case, 
the vortex is thicker 
on the equatorial region when the
black hole has charge. This effect is of order $p\sim 1/N$, 
and acts against the 
``squeezing'' of the string due to the black hole attraction. Intuitively,
the presence of
charge induces tension ---a repulsive effect.

However, there is a more important modification introduced by a non-zero 
charge on the black hole. 
If we compute the magnetic flux crossing any portion of
the horizon, which, from (\ref{pfield}), is given by 
\begin{equation}\label{fsol}
{}F_{\theta\varphi}|_{r=r_+} = - p(r_+^2 - Q^2)\sin2\theta,
\end{equation} 
we see that it decreases as we increase the charge, until
it precisely vanishes for an extreme black hole. 
Moreover, we see from (\ref{bsol}) that the Higgs field also vanishes
at the horizon in that limit. The extreme black hole expels from its
horizon {\it all} the fields that live in the core of the string!

It was already known that an extreme black hole placed in a  
uniform magnetic
field exhibits a sort of ``perfect diamagnetism'' (see \cite{cham}
for a review of the literature and an extended analysis of this effect). 
The solution 
(\ref{pfield}) for the gauge field describes precisely this effect. 
But here we have found 
that this exclusion is also true for the Higgs field associated to 
the string vortex.
Moreover ---and this is something that we could not have anticipated 
from what we knew about the behavior of the magnetic fields--- a 
global string is also expelled
from the extreme horizon. This is very easy to see: simply set $p=0$ in 
(\ref{bsol}) to obtain the field of the global string.

Given that the solution we have found is only a leading order
approximation for large $N$, one might inquire whether further 
corrections still preserve the expulsion of the fields. The numerical
evidence from next section confirms this point, even down to $N=1$.

A natural question to ask is whether the black hole will stay inside 
the vortex, or will instead try to find its way outside the core. 
To this effect we will study the energy stored in the string core 
when a black hole is sitting inside it. 

{}For a static solution of the Abelian Higgs equations, in 
length and energy units rescaled by the Higgs wavelength, 
the energy density takes the form
\begin{equation}\label{engdens}
T_0^0 = -g^{ij}\partial_i X \partial_j X - X^2 g^{ij} P_i P_j +
{\beta \over 2} F^2 + {1\over 4} (X^2-1)^2.
\end{equation}
More specifically, for the vortex in the Reissner-Nordstrom background,
\begin{eqnarray}\label{rnengdens}
T_0^0 &=& V(\partial_r X)^2 + {1\over r^2} (\partial_\theta X)^2 + 
{N^2 X^2 P^2\over r^2\sin^2\theta} + {1\over 4} (X^2-1)^2 \nonumber\\
&+&
{N^2\beta \over r^2\sin^2\theta} \left[ V(\partial_r P)^2 + {1\over r^2} 
(\partial_\theta P)^2 \right].
\end{eqnarray}

Let ${\cal E}_{0,bh}^{(g)}$ be the total energy of the 
gauge field in the absence or 
presence of the black hole. A rather long
analysis, which we will spare the reader, keeping leading order terms in $1/N$ 
(or, what amounts 
to the same, expanding for small $p$) leads to the conclusion that
\begin{equation}\label{engratio}
{{\cal E}_{bh}^{(g)}\over {\cal E}_0^{(g)}} = 1 - c E\sqrt{p} +O(p),
\end{equation}
where $c$ is a positive constant (a pure number) of order unity. 
Hence, the presence of a black hole within
the vortex decreases the energy of the gauge field. For fixed black hole
mass $E$ this is independent of the value of the charge. Even if the
latter causes an equatorial thickening of the string 
which would tend to increase the energy, the energy of the fields
decreases.  This is, however, a smaller effect 
of order $O(p)$. 

Consider now the energy stored in the Higgs field. 
{}From the solution (\ref{bsol}) we can see that switching on a 
black hole mass 
$E$ decreases the value of the Higgs field inside the vortex. Again, the 
charge works in the opposite direction, but this is a smaller effect. 
As regards the energy, the largest contribution is
the potential energy arising from
the fact that the core is in the false vacuum. This is, 
however, hardly affected by the introduction of the black hole. The gradient
terms, on the other hand, are more significantly modified, and it is not 
difficult to see that a non-zero value of the black hole mass $E$ always 
tends to decrease the energy.

Of course, these energetic considerations alone do not tell us what the 
forces induced by the vortex on the black hole are. 
In order to compute these, we 
would need to consider configurations where the black hole is not 
exactly at the axis of the vortex, and thus, non-axisymmetric 
configurations. 
A simple way to estimate the forces would be to
compute the energy stored in the vortex as a function 
of the separation $x$
of the center of the black hole to the axis of the vortex, call
this function ${\cal E}(x)$. 
It is clear that the lack of symmetry makes this 
problem very much harder. Nonetheless, the estimations above give us
the values ${\cal E}(x=0)={\cal E}_{bh}<{\cal E}(x\rightarrow \infty)= 
{\cal E}_{0}$.
If ${\cal E}(x)$ were a monotonic function of $x$, which does not seem 
unreasonable, then the forces acting
on the black hole would tend to keep it inside the vortex.

The conclusion seems to be that the black hole should remain
stable inside a thick vortex. Qualitatively, this is largely 
independent of the presence
or absence of black hole charge, and, in particular, of the vanishing of 
the fields
on an extreme horizon. However, as will be revealed in next section, 
this no longer remains true if the vortex radius
shrinks below the horizon radius.

\section{Numerical solutions}

In the previous section we have provided evidence that an extremal black hole
does not allow  penetration through its horizon of the fields associated with
the vortex. The analysis, though, has had to be restricted to the situation
where the black hole is small relative to the vortex and stays 
well within its core. It is irresistible to push this picture 
to its limits and let the vortex shrink to a size smaller
than the horizon radius. Will the string still fail to pierce the extreme
horizon? In this case we would expect that the presence of the black hole 
inside the vortex should cause an increment of the tension of the 
flux. As a result, the energy stored in the vortex should increase 
--instead of decrease, as in the previous section-- and this would 
clearly suggest that the configuration is unstable: 
the extremal black hole would
strongly oppose wearing the Abelian Higgs wig, 
and the (thin) string should slide off 
the horizon, leaving the extreme black hole as bald as we have always known it to be.

To analyze these issues we shall need to resort to numerical integration of
the equations (\ref{rnoeq1}) and (\ref{rnoeq2}) outside and on the black hole
horizon. Our results will 
confirm the picture of the previous section for thick vortices,
as well as provide evidence that, when the
string is thin, it will tend to slip off the extreme horizon.

The Abelian Higgs equations in the presence of a background
Reissner-Nordstrom metric are elliptic. On the horizon they become
parabolic. In order to solve the equations numerically, we use a
technique first used by Ach\'ucarro, Gregory and Kuijken \cite{ana}. 
We will briefly describe this technique below.

One common approach to solving elliptic equations is to introduce an
artificial, first-order in time, diffusive term to the elliptic
equation to be solved. The resulting diffusion equation is then
iterated, and the fields relaxed, 
until the time dependent term (the `residual') approaches zero to sufficient 
accuracy, leaving a
solution to the original elliptic equation. This is the basic
technique used in \cite{ana}, however, they have introduced some changes
in order to solve the equations on the horizon.

Their method is to set boundary conditions at $\theta = 0$ and $\theta
= \pi$ consistent with field values at an Abelian string core. At $r =
\infty$ boundary conditions are set to those of the asymptotic fields
of the string. Field values on the horizon are also initially set to
asymptotic values. The integration technique then proceeds as follows:

First, the discretised field is relaxed inside the
simulation volume. Next, using the equations for the fields on the
horizon, which are elliptic in the radial direction, the field is
relaxed on the horizon, giving new boundary points there. This process
is iterated until the residual is considered small enough that a
solution has been found.

The relaxation procedure we have used is based on the successive
over-relaxation method described in \cite{num}. However,
since the equations are nonlinear, Chebyshev acceleration had to be
turned off, and we typically had to under-relax the field.

To check our code, we ensured that solutions to the uncharged black
hole case matched those of \cite{ana}.

The discretised equations for the $P$ and $X$ fields in a Reissner-Nordstrom
background outside the horizon are

\begin{small}
\begin{eqnarray}
X_{00} = \frac{\frac{2}{r} (1 - \frac{E}{r}) \frac{X_{+0} - X_{-0}}{2 \Delta
r} + \frac{\cot{\theta}}{r^2} \frac{X_{0+} - X_{0-}}{2 \Delta \theta}
+ (1 - \frac{2E}{r} + \frac{Q^2}{r^2}) \frac{X_{+0} + X_{-0}}{\Delta
r^2} + \frac{X_{0+} + X_{0-}}{r^2 \Delta\theta^2}}
{(1 - \frac{2E}{r} + \frac{Q^2}{r^2}) \frac{2}{\Delta r^2}
 + \frac{2}{r^2 \Delta\theta^2} + \frac{1}{2} (X_{00}^2 - 1) +
 (\frac{N P_{00}}{r \sin\theta})^2 }
\end{eqnarray}
\end{small}
\begin{small}
\begin{eqnarray}
P_{00} = \frac{\frac{2}{r^2} (E - \frac{Q^2}{r}) \frac{P_{+0} -
 P_{-0}}{2\Delta r}  - \cot{\theta} \frac{P_{0+} - P_{0-}}{2 r^2
 \Delta\theta}
 + (1 - \frac{2E}{r} + \frac{Q^2}{r^2}) \frac{P_{+0} + P_{-0}}
{\Delta r^2}
 + \frac{P_{0+} + P_{0-}}{r^2 \Delta \theta^2}}
{(1 - \frac{2E}{r} + \frac{Q^2}{r^2}) \frac{2}{\Delta r^2} +
\frac{2}{r^2 \Delta \theta^2} + \frac{X_{00}^2}{\beta}}
\end{eqnarray}
\end{small}
and the $P$ and $X$ equations on the horizon are
\begin{equation}
X_{00} = \frac{\sqrt{E^2 - Q^2} \frac{X_{+0}}{\Delta r} + \frac{X_{0+}
+ X_{0-}}{2 \Delta\theta^2} + \cot\theta \frac{X_{0+} - X_{0-}}{4
\Delta\theta}}
{\frac{\sqrt{E^2 - Q^2}}{\Delta r} + \frac{1}{\Delta\theta^2} +
\frac{r_+^2}{4} (X_{00}^2 - 1) + \frac{1}{2} (\frac{N
P_{00}}{\sin\theta})^2} 
\end{equation}
\begin{equation}
P_{00} = \frac{\sqrt{E^2 - Q^2} \frac{P_{+0}}{\Delta r} + \frac{P_{0+}
+ P_{0-}}{2 \Delta\theta^2} - \cot\theta \frac{P_{0+} - P_{0-}}{4
\Delta\theta}}
{\frac{\sqrt{E^2 - Q^2}}{\Delta r} + \frac{1}{\Delta\theta^2} +
\frac{r_+^2}{2\beta} X_{00}^2}
\end{equation}

Here, a zero subscript indicates the value at a given meshpoint
and $+$ and $-$ indicate adjacent values to the left or right.

On the $\theta = 0, \pi$ boundaries we set $P = 1$ and $X = 0$, at $r
= r_{max}$ we set $P = 0$ and $X = 1$, and initially on the horizon,
we set $P = 0$ and $X = 1$. The boundary conditions at $r = r_{max}$
are only an approximation to the correct values since the string is
forced to have a width of one gridzone at $r_{max}$. This tends to
distort the field values near $r_{max}$. In our simulations, we have
solved the equations on a cartesian $r$-$\theta$ mesh. In order to
minimize the distortion, we set $r_{max}$ to be from $5$ to $10$
horizon radii. Since gridzone volume increases for large $r$, the
string is then well approximated as having a width of less than a
gridzone in $r-{\theta}$ coordinates.

With the above discussion in mind, we now present the numerical results.

\subsection{Expulsion of the electromagnetic and Higgs fields by the extreme
black hole}

We have already seen, in equations (\ref{bsol})-(\ref{fsol}), that when the
vortex size is large compared to the black hole size the magnetic and Higgs
fields are both expelled by the extreme black hole.  However, the estimates
which we used to obtain these analytic expressions no longer hold when the
vortex is very skinny relative to the hole.  In this situation, we have to
use numerical techniques.

We have pushed this calculation to the limits, making the vortex as small
as we could given the computational constraints.  What we have found is that
the vortex is {\it always} expelled, no matter how small the magnetic and Higgs
flux tubes are taken to be.

Here we present dramatic pictures of the numerical evidence which we have 
amassed.  Our intention is to give the reader a `flavour' of the general
phenomena using a frugal selection of images.  The general pattern displayed
here holds no matter how small you make the flux tubes.

We begin with the expulsion of the $P$ field by the extreme hole.  In the
diagram below, we have set $E = Q = 10$, with winding number $N = 1$ (the smallest
winding possible). Furthermore, the Bogomolnyi parameter $\beta$ is set 
equal to unity, so that the magnetic and Higgs flux tubes are the same size:\\
\vspace*{0.3cm}

\epsfxsize=6cm
\epsfysize=12cm
\hspace*{3.5cm} \epsfbox{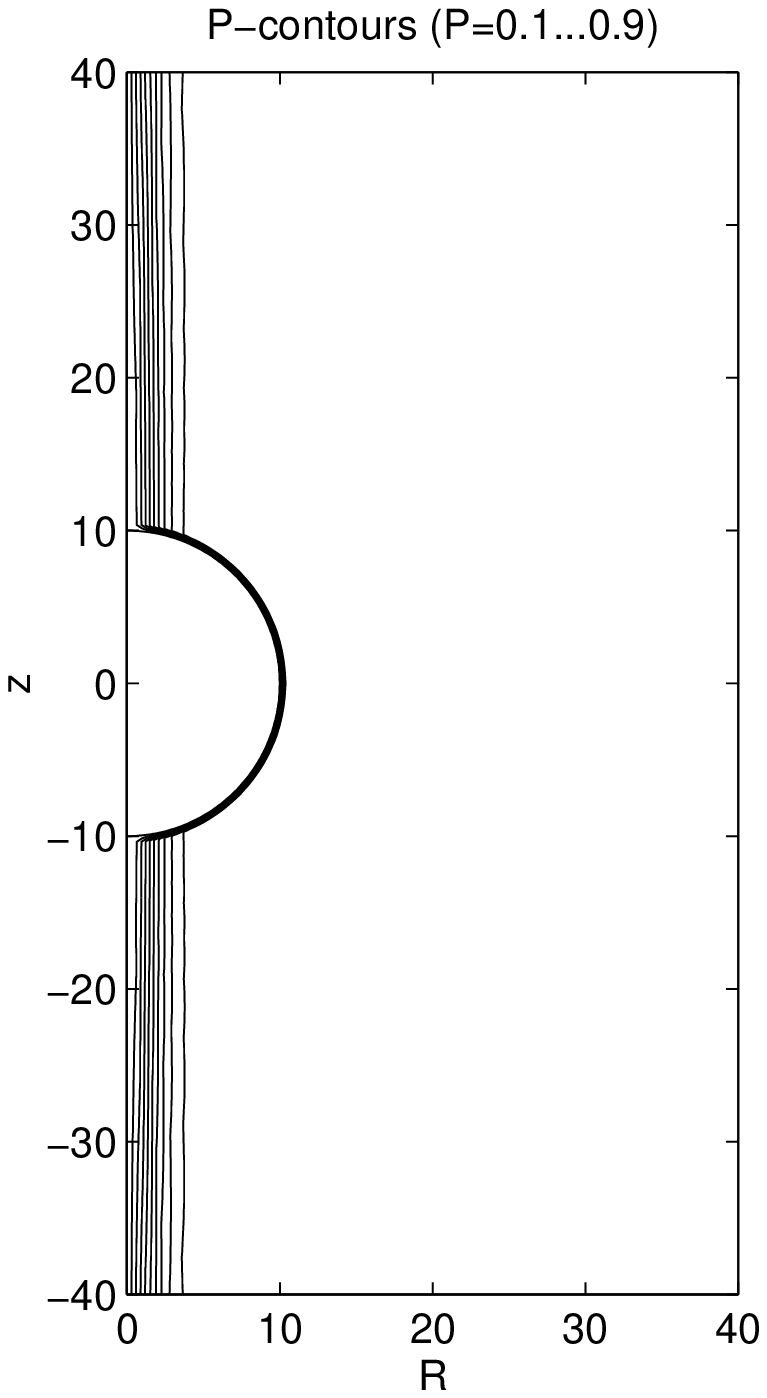}
\vspace*{0.7cm}

Clearly, the $P$ field literally `wraps' the black hole horizon; furthermore,
given the relation between $P$ and $F_{{\theta}{\varphi}}$ it is clear that no
magnetic flux is crossing the horizon.  The extreme hole still behaves
just like a perfect diamagnet.  We now want to see if we can `puncture' the
horizon with flux by making the magnetic flux tube even smaller.  Perhaps
the simplest way to make the vector flux tube thinner is by decreasing the
value of $\beta$. This has the effect of greatly
enhancing the size of the mass term in (\ref{rnoeq2}). 
Since $\beta$ is the ratio of the sizes of the vector and
Higgs flux tubes, making $\beta$ very small will correspond to making 
the magnetic flux tube very skinny (while keeping the size of the 
Higgs flux tube fixed, though enlarging the size of the transition 
region between massless and massive Higgs phases, which is 
$\sim \beta^{-1/2}$).  
This is done in the diagram below, where we have set
$E = Q = 10$, $N = 1$, and ${\beta} = .0001$:\\
\vspace*{0.3cm}

\epsfxsize=6cm
\epsfysize=12cm
\hspace*{3.5cm} \epsfbox{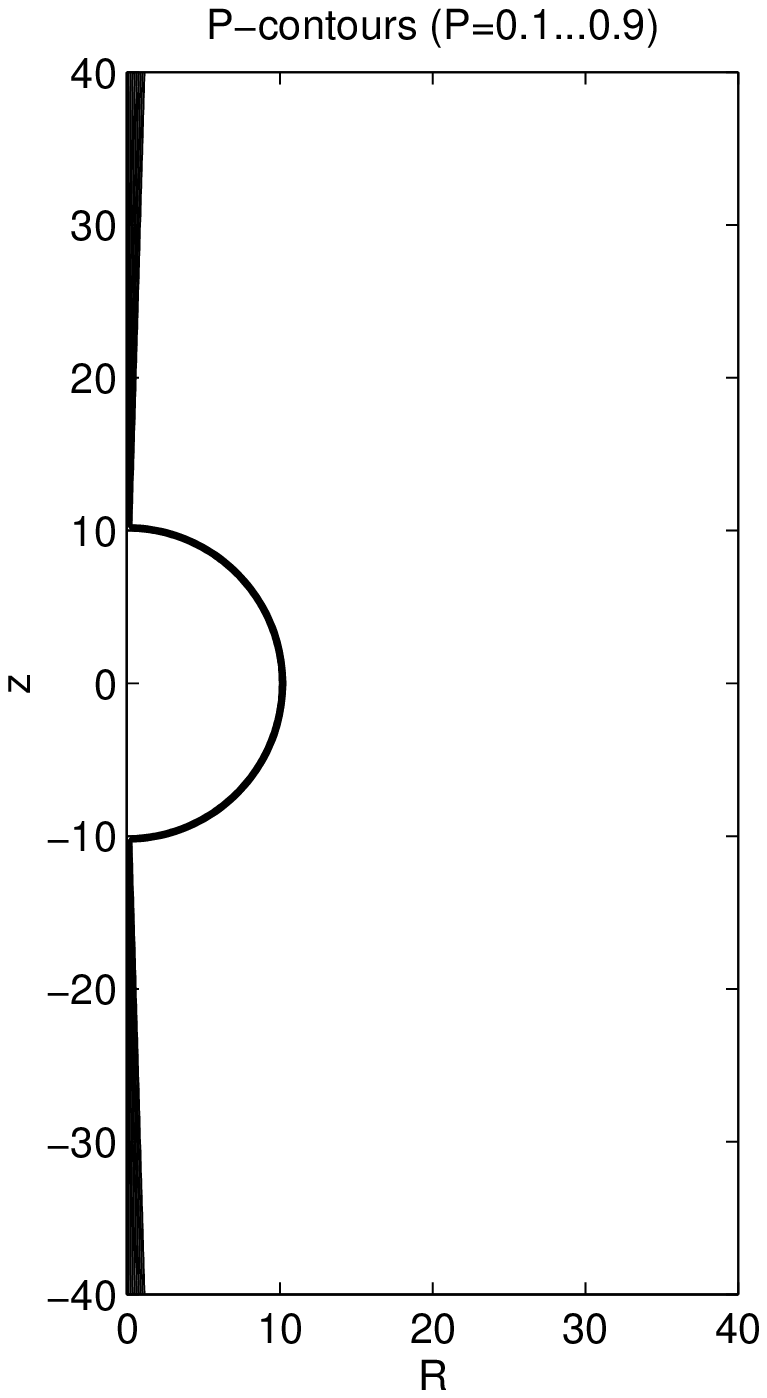}
\vspace*{0.7cm}

Again, the $P$ contours all wrap around the black hole horizon, indicating
that there is never any penetration.  We have repeated this calculation for
the smallest resolvable values of $\beta$ (keeping $E$, $Q$ and $N$ fixed) and
we have always seen the same phenomena.  Similarly, we have kept $N$ and 
$\beta$ fixed and made $E = Q$ very large (i.e., fixed the vortex 
size and increased the
size of the extreme hole) and again we see perfect expulsion.  
(There is very little reason to show the pictures of these calculations 
since without magnification they are qualitatively identical to the figure 
above.)

We now turn to the behaviour of the Higgs field $X$.  Again, in the thin
vortex limit we are unable to make analytic estimates and we are forced to 
resort to numerical integration.  We have found that the $X$ field is always
expelled from the extreme hole, no matter how small the scalar flux tube is
made.  Actually, in the figures below what we do is fix the
size of the scalar flux tube (by fixing $N = 1$ and ${\beta} = 0.5$) and we
allow the mass of the extreme hole to increase.  The plots run from left to right
with increasing mass.  The graphs are plotted for the values $E = Q = 5$,
$E = Q = 10$, $E = Q = 20$, and $E = Q = 30$.
\vspace*{0.6cm}

%\epsfxsize=2.5cm
%\epsfysize=9cm
%\hspace*{3.5cm} \epsfbox{pig-fig.ps}
\centerline{\epsfxsize=4cm \epsfysize=8cm \epsfbox{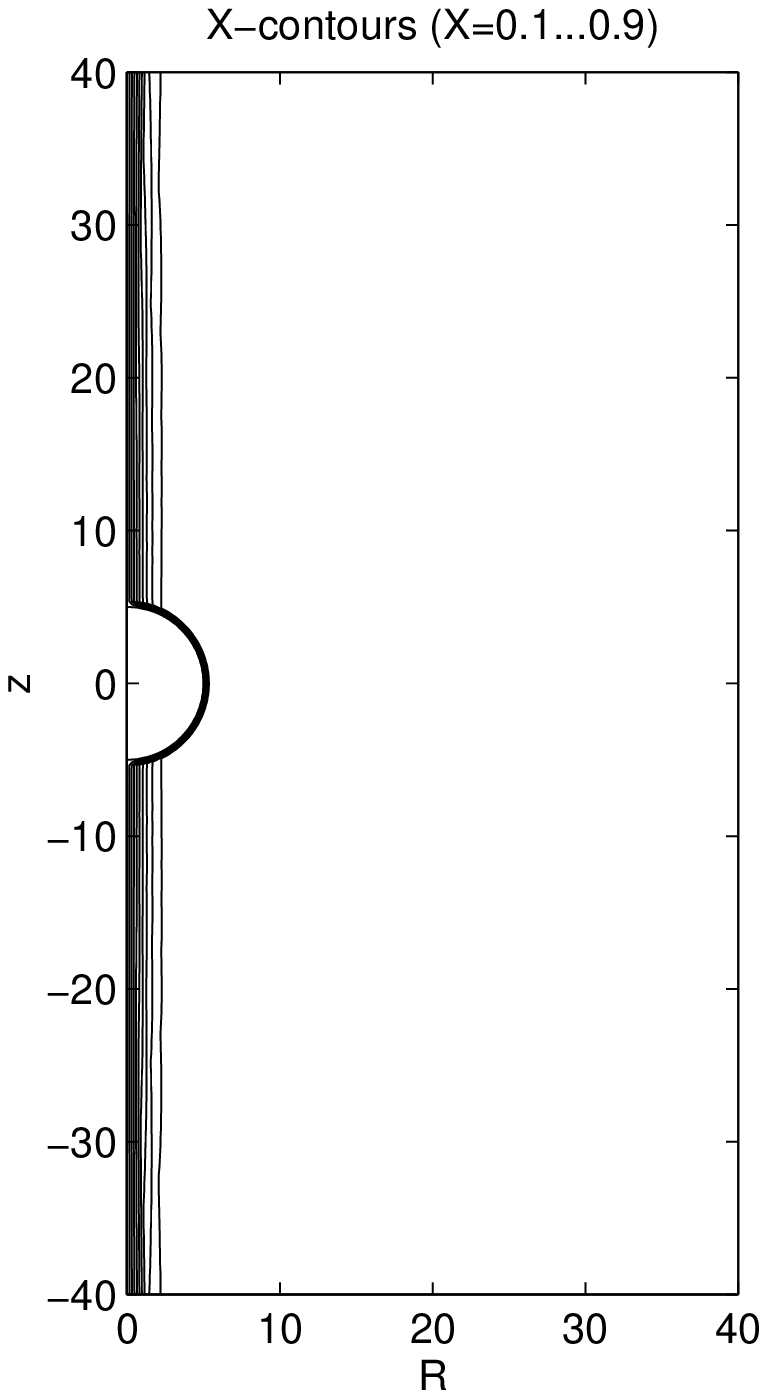}  
\epsfysize=8cm \epsfxsize=4cm 
\epsfbox{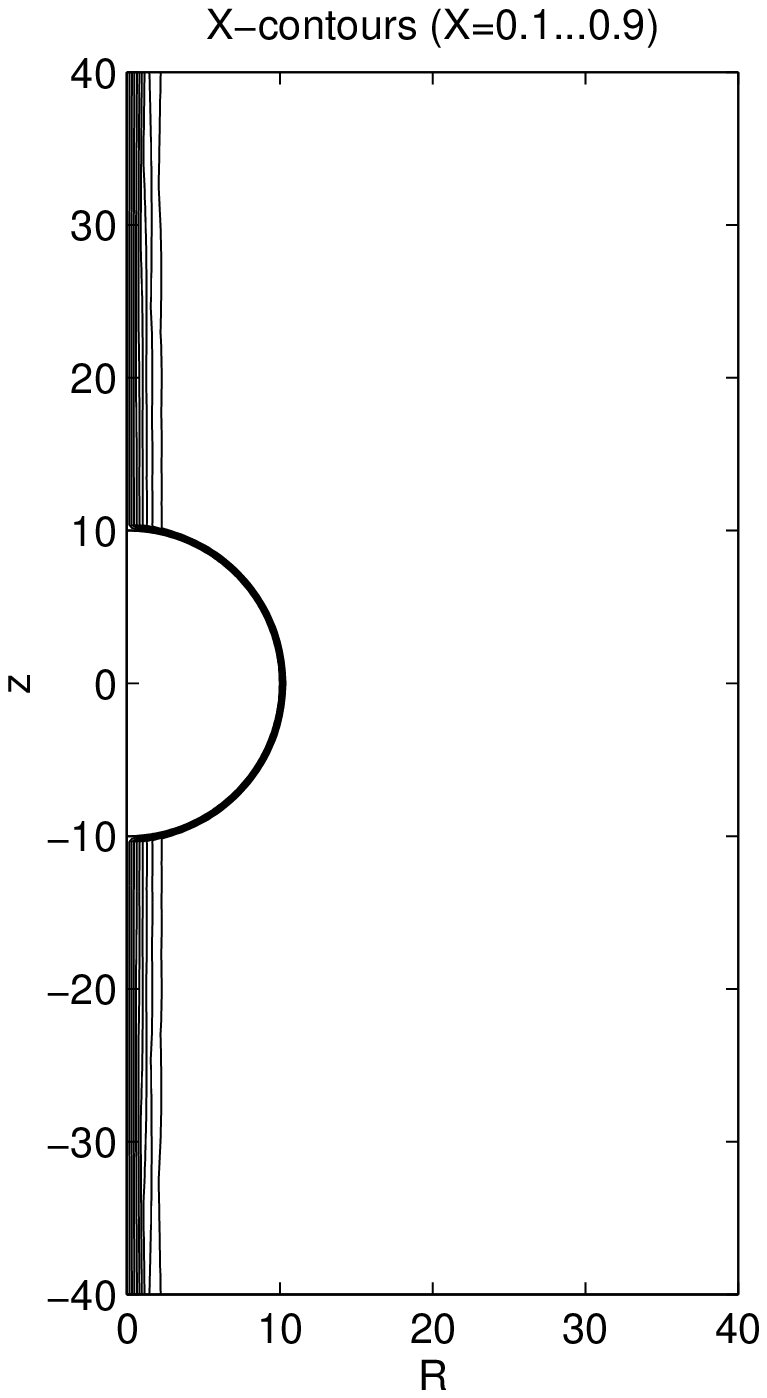} \epsfysize=8cm \epsfxsize=4cm \epsfbox{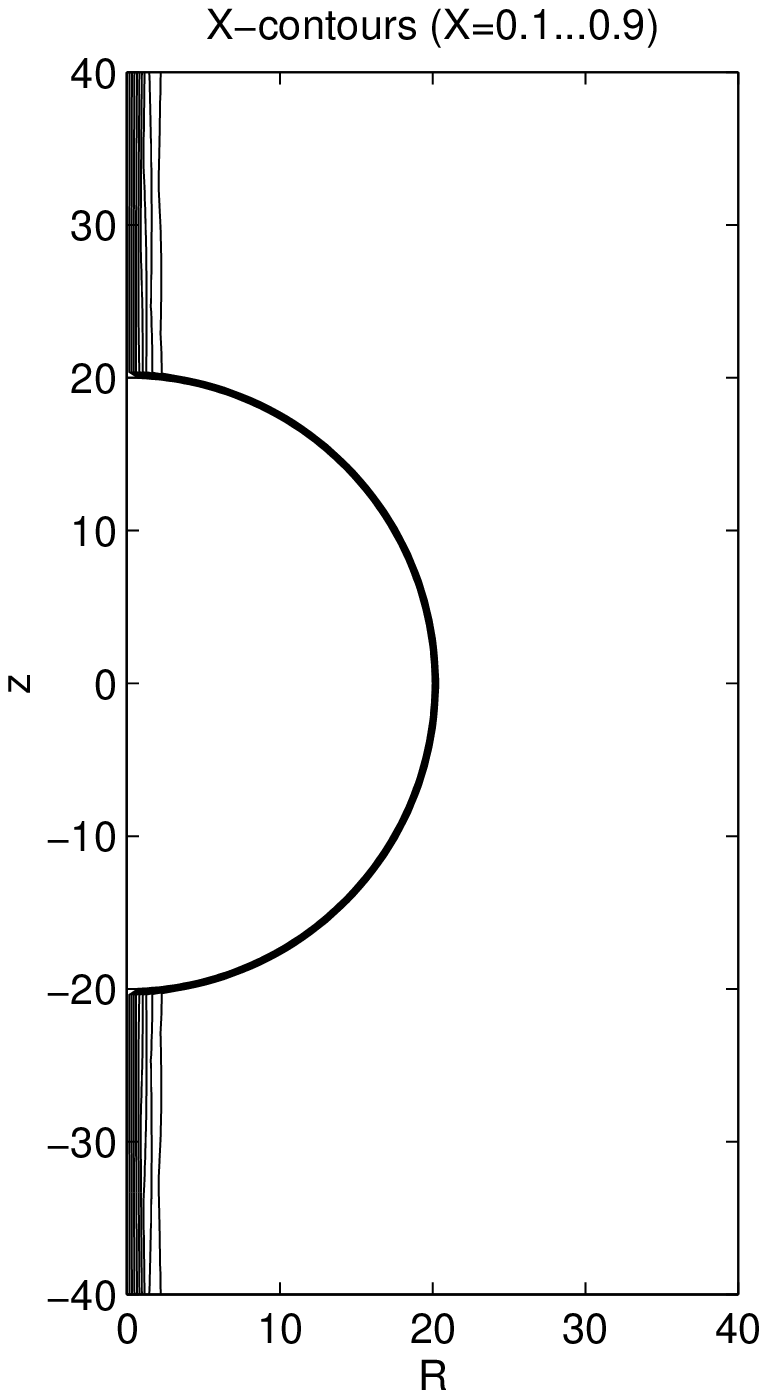}
\epsfysize=8cm \epsfxsize=4cm \epsfbox{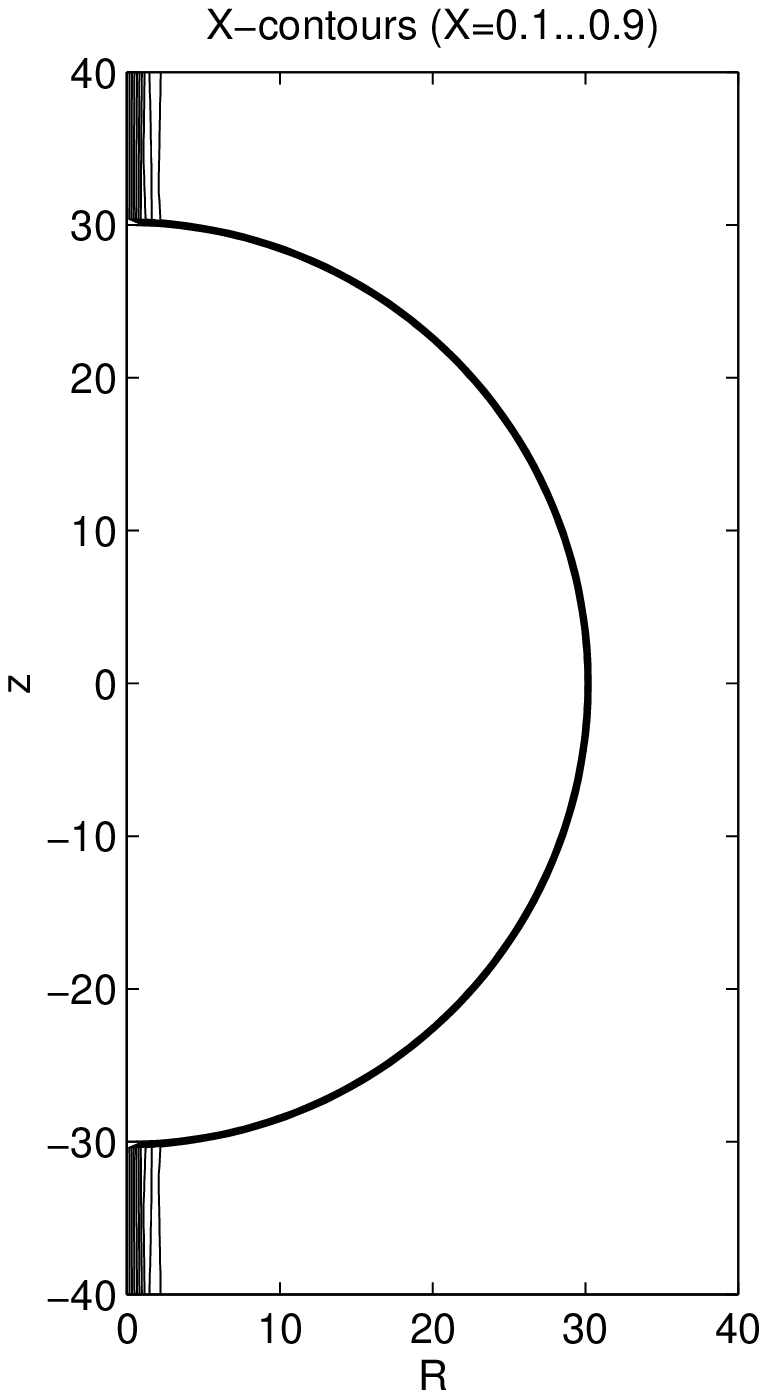}}
\vspace*{0.7cm}

As was claimed, the $X$ contours all wrap around the black hole horizon,
no matter how large the hole is made.  Indeed, the above sequence of pictures
provides an intuitive picture of why equation \ref{engratio} makes sense.
When the black hole is much smaller than the vortex, the black hole is just a
`hole' where no vortex energy can be stored.  Thus, the presence of the hole
tends to subtract the total energy of the vortex.  On the other hand, when
the hole becomes much larger than the vortex (and our estimates break down),
the vortex still has to wrap the hole and so we would expect the total energy
of the vortex to become very large.  We now provide a more detailed discussion
which will show that this intuition is in fact correct.

\subsection{Instability of the vortex energy in the large mass limit}

As we have discussed, eq. \ref{engratio} tells us that when the hole is
small relative to the vortex increasing the mass of the hole tends to decrease
the total energy stored in the vortex.  We can also see it must be the
case that when the hole is very large relative to the vortex, increasing the
mass of the hole must increase the energy of the vortex
due to the tension in the flux lines.  Thus, the energy
of the vortex as a function of extreme black hole mass must have at least one
minimum.  In fact, it is not hard to see that there must be at most one minimum
(although we will not provide an analytic argument here, since the numerical
results will make this clear).  We shall denote this value of the hole mass,
where the vortex energy is minimized, as $E_c(N, {\beta})$.  We have written
$E_c$ as a function of $N$ and $\beta$ in order to emphasize that the critical
mass depends on the `width' of the vortex.  Now, again let ${\cal E}_{bh}$
denote the total energy of the vortex centered on the extreme black hole 
(note: in the numerical calculations which follow we have introduced an obvious
cutoff, i.e., we do not integrate over all of spacetime to obtain the energy, 
rather we integrate out to the boundaries of some large `box').  Then it is 
always the case that ${\cal E}_{bh}(E_c) < {\cal E}_{0}$, where ${\cal E}_{0}$
is the energy of the vortex in the absence of the black hole.  This means
that a black hole of mass $E_{c}$ is perfectly happy to sit inside of the 
vortex, and indeed it would be energetically unfavourable for the hole to be
removed from the vortex.  In fact, it is always the case that there exists a 
maximum mass $E_{max}$ such that for all black holes of mass $E < E_{max}$,
${\cal E}_{bh}(E) < {\cal E}_{0}$; as long as the hole is not too massive, 
it is content to sit inside the vortex. 

The statements made above are based on the results of our numerical computations
of the total energy ${\cal E}_{bh}$.  In the figure below we have plotted
the results of one such computation.  Here, we have set ${\beta} = 0.5$ and
$N = 10$.  The flat, horizontal line (at 6640) represents ${\cal E}_{0}$
in our units.  
Clearly, for these values $E_c$ is about 8 and $E_{max}$ is about 15.
Furthermore, for black holes of mass greater than $E_{max}$ the energy of the
vortex is diverging.  The erratic behaviour of the vortex energy for very
small values of the black hole mass is an artifact of the numerical techniques
employed in the calculation and should be ignored:\\
\vspace*{0.3cm}

\epsfxsize=11cm
\epsfysize=8cm
\hspace*{2cm} \epsfbox{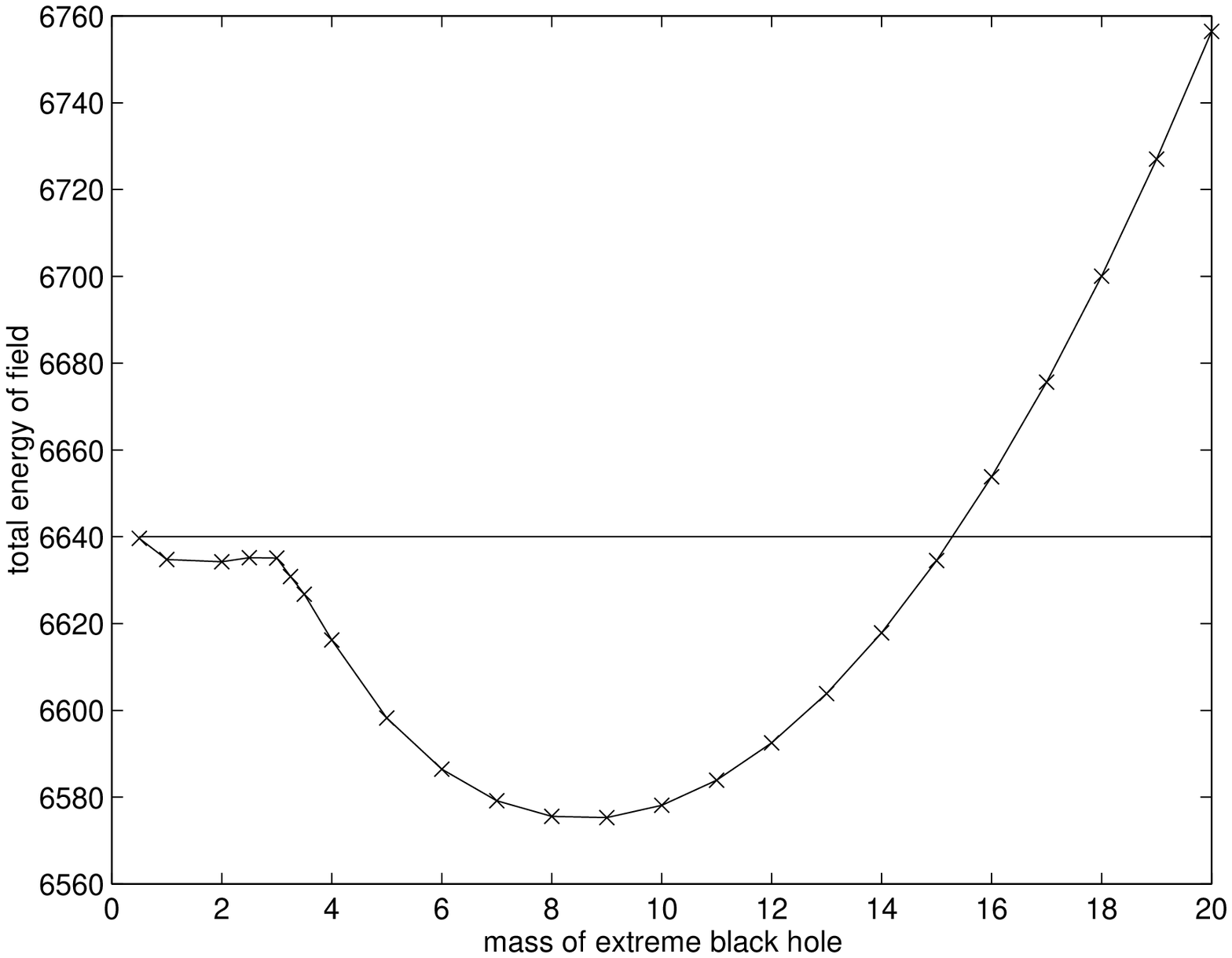}
\vspace*{0.7cm}

It is clear from this graph that a black hole with mass $E > 15$ is going to
find it energetically favourable to slip out of the vortex.  Thus, it is really
not appropriate to think of such a vortex as a `property of the black hole';
the identification of the vortex as long hair does not seem to go through
in this situation.  Of course, when the mass of the hole is small you could
still technically try to identify the vortex with hair since at least in that
case the configuration is energetically stable.  On the other hand, the fact
remains that the vortex is completely expelled from the hole, even in the 
(putatively) stable situation. Thus one would say that the vortex is not 
{\it dressing} the black hole. It is still unclear to us whether or not 
one should think of such a `thick' vortex as genuine hair for a small 
extreme black hole. This is somewhat a reversal of previosly studied 
situations (e.g., the coloured black holes), where the black hole may 
be dressed but the configuration is unstable.

\subsection{No expulsion of vortex fields in the near extreme limit}

So far we have presented firm numerical evidence that the fields of an
Abelian Higgs vortex are expelled from the horizon of an extreme black hole.
A natural question is then whether or not similar results continue to hold
when the hole is made {\it slightly} non-extreme.  As is well known, a 
non-extreme black hole with non-singular horizon has $Q < E$.  As we let
$Q$ approach $E$ from below (letting the hole approach extremality) will
we see the fields $P$ and $X$ `gradually' expelled from the horizon?  Or will
the fields suddenly `pop' out only when we get precisely to the extreme limit?

In order to understand how to answer this question, it is useful to first 
recall the estimates which we made in Section II in the limit where the 
vortex is thick compared to the outer horizon radius of the black hole.  
In particular, recall equation \ref{fsol}, which follows immediately 
from equation \ref{pfield}.
Equation \ref{fsol} tells us that, in regions where the mass of the gauge field
is negligible, the magnetic flux across the horizon
in the non-extreme limit will always be non-vanishing, and hence that the 
vortex $P$ field will penetrate the horizon.  The flux vanishes in the extreme
limit since the equation says that $F_{{\theta}{\varphi}}$ on the horizon
is proportional to ${r_{+}}^2 - Q^2$ (where $r_{+}$ is the outer horizon radius) 
and $r_{+} = Q$ precisely in the extreme limit.

Now, the region $r_{-} < r < r_{+}$ is regular in the coordinates which we
have been using, even though there are (removable) coordinate singularities
at the inner and outer horizons $r_{-}$ and $r_{+}$.  Furthermore, there
is no reason why the estimate \ref{fsol} shouldn't continue to hold in this
region.  In other words, there is always a surface at $r = Q$, which we 
dub the ``Meissner surface'', across which no flux may flow.  This Meissner
surface agrees with the outer black hole horizon only in the extreme limit,
and so it is only in this limit that the Meissner surface is of relevance
to external observers. One could think that, since for a near-extremal 
black hole the Meissner surface
can be very close to the (outer) horizon, then if the layer of vortex on the 
Meissner surface is thick enough, the expulsion from the Meissner surface 
might be appreciable by external observers. Now, this vortex layer gets 
thicker with vortex size. But for large vortices the effect of the 
Meissner surface can be read from (\ref{pfield}), and we see that the 
expulsion only appears when the extremal limit is reached.

In all of our numerical calculations, we do not consider the penetralia of the
black hole.  Rather, we solve for test fields outside (and on) the horizon of
the hole and we do not concern ourselves with what goes on inside the horizon.
This is why, {\it by construction}, we do not expect to see the fields
gradually expelled from the horizon.

For the edification of the reader we present here some pictures of calculations
which show that the argument given above goes through even when the vortex is
thin relative to the radius of the hole.  In the below calculation, where we 
plot $P$, we have set $E = 10$, $Q = 9.99$, $N = 1$ and ${\beta} = 1$:\\
\vspace*{0.3cm}

\epsfxsize=6cm
\epsfysize=12cm
\hspace*{3.5cm} \epsfbox{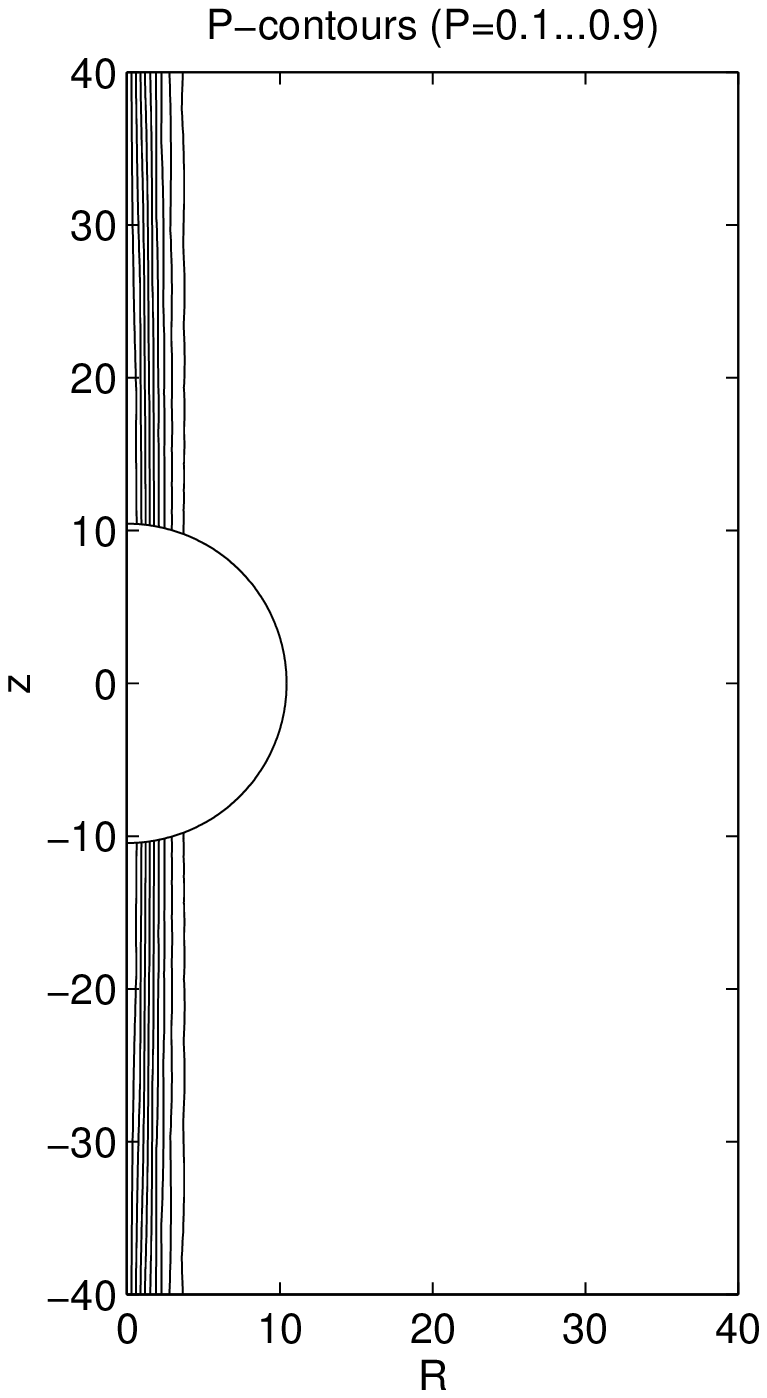}
\vspace*{0.7cm}

Clearly, the $P$ field is passing right through the black hole horizon even 
though the hole is quite close to extremality.  Similarly, one finds that the
$X$ field contours flow through the outer horizon of any non-extreme black hole.
In the below calculation, where we plot $X$, we have again set
$E = 10$, $Q = 9.99$, $N = 1$ and ${\beta} = 1$:\\
\vspace*{0.3cm}

\epsfxsize=6cm
\epsfysize=12cm
\hspace*{3.5cm} \epsfbox{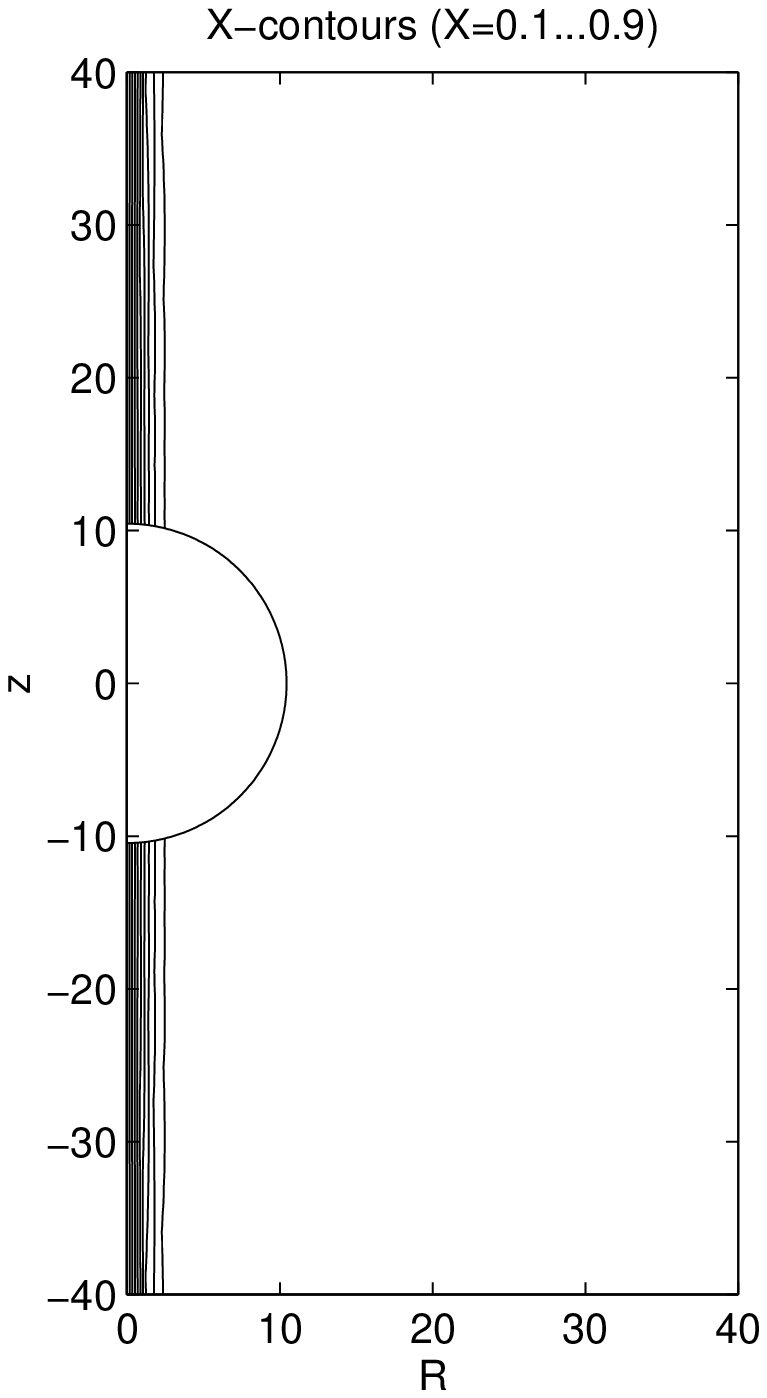}
\vspace*{0.7cm}

\section{Conclusions}

We have found that the fields of a vortex are
always expelled from an extreme horizon; this
effect is generic for
arbitrary relative sizes of the horizon and the vortex core. Furthermore, a 
thin enough vortex tends to slip off the black hole. Thus, it appears 
that an extreme black hole cannot support `long' abelian Higgs hair.
Of course, we have in no way accounted for the
back reaction of the vortex on the geometry.
Is there any reason why the flux tubes {\it shouldn't}
pierce the horizon once back reaction is included?  
Actually, there is a piece of evidence that the expulsion may hold 
exactly: there do exist {\it exact} solutions  (i.e., including the full back 
reaction) for black holes in $U(1)^2$ theories where a black hole that is
charged to extremality with respect to one of the gauge fields, completely
expels the field of magnetic configurations of the other 
gauge field \cite{cham}.
In these solutions none of the gauge symmetries is broken, but recall that
the spontaneous symmetry breaking is of negligible influence on the 
perturbative first order solution inside the core that we have found 
in (\ref{pfield}). This strongly suggests that, after 
accounting for the backreaction, the flux should be expelled from a black
hole that sits inside it, at least in the case where the vortex is thick. 
In view of the evidence provided above, the 
effect could as well persist for thinner black holes, but we cannot be 
conclusive. In any case, the back reaction would certainly be expected to be
small if the energy scale of symmetry breaking is sufficiently small compared 
to the black hole mass.

In order to implement back reaction in the numerical calculations, 
we would first start with a fixed background and solve
for the `test fields' as we have done in this paper. Next, we would have to
plug the energy-momentum tensor for the test fields into the Einstein equations
and solve for the `corrected' background geometry.  Then we would again have
to solve for the vortex field configuration in the corrected geometry, and so 
on. Now, in general the horizon will move each time we
obtain a corrected background geometry. While we are currently working on a 
numerical approach to include the back reaction, we will have nothing more 
to say about this issue here.

We have argued that vortices fail to penetrate 
extreme horizons. Will this still hold true if the string tries to 
{\it end} at a black hole?
It has been argued in \cite{ana,ehkt} that there is no global topological
obstruction for a topologically stable string to end at a black hole. The 
reason is that the spatial topology $S^2\times {\bf R}$
of the extended black hole spacetime allows one to take gauge patches \`a 
la Wu-Yang that transform the trivial vacuum on one side of the black hole 
to the non-trivial configuration of the vortex on the other side. 
Furthermore, there do exist solutions where the string 
actually penetrates the non-extreme black hole, as shown for Schwarzschild 
in \cite{ana} and generalized to Reissner-Nordstrom in this paper. 
Now imagine an open string ending on a non-extreme charged horizon. Although
the string will want to pull the black hole, and then accelerate it, we need
not consider this effect for as long as we neglect the gravitational 
backreaction of the string; in other words, the black hole is too massive 
for the string tension to accelerate it in any appreciable way. Therefore
it is consistent with our approximations to treat the system as a static one.
Then, increase the charge of the black hole until the extremal limit is 
reached, just as we have done for the string threading the black hole. In the
latter case the string is expelled at extremality. But in the situation we
are considering now there is a crucial difference. A topologically stable
string cannot have endpoints, and therefore when extremality is reached
the open string cannot wrap the horizon, nor simply detach from it. It would
appear that the string should remain attached to the horizon. 
This conclusion, if correct, is rather
puzzling. It is not clear from the local field equations, which
determine the expulsion of the field, why the string could end at, but not
thread, the black hole. 
As a matter of fact, one would tend to think that if a string can end 
at a black hole, then it should be able to thread it as well! The reason is
simple: 
a string threading the black hole 
could be constructed by the simple procedure of 
attaching to the horizon the endpoints of two strings, carrying ingoing and 
outgoing fluxes. 
But configurations threading extremal horizons are ruled out by the
evidence presented in this paper.

Thus we find a sharp problem here, which remains an interesting extension 
of our work. There is, in fact, 
one line of research where all this finds direct application. 
There have been a number of papers describing the pair creation of 
black holes with strings ending on them \cite{haro,emp,ehkt,grhi,emp2}. 
Apart from the topological stability issues, the process of a string
snapping with formation of black holes differs in one important 
respect from the strings that break with monopoles at the end.
In order for the Euclidean gravitational instanton that 
mediates the process to be regular, the black holes must have (unconfined) 
charge, and be either extremal or close to extremality. 
This forces one to introduce, in addition to the massive gauge field
carried by the string, a (massless) $U(1)$ field to which the black hole
charge couples. Effectively, one works in a $U(1)^2$
theory of the same kind we have been discussing in this paper. 
The question of whether the string can or not end at the extreme horizon
must necessarily be addressed prior to the construction of
the corresponding instanton. The arguments above suggest that this may
not be an easy question.
We see that consideration of `realistic' strings must be addressed to 
determine whether new selection rules, of a 
sort somewhat different from those recently discussed in \cite{ag}, might
be at work.

Another interesting scenario involving pair creation of black holes, 
still in a theory with two gauge fields 
${\cal F}$ (massless) and $F$ (massive), is the following: let there be a 
string vortex (carrying confined flux of $F$) 
and a magnetic (unconfined) background field ${\cal B}$ parallel to the 
string. Suppose that a pair of magnetic
holes, charged relative to the ${\cal B}$ field with charges ${\pm q}$ 
are pair created and accelerate apart under the 
force induced by the field, like in the Schwinger pair creation process. 
Suppose, moreover, that the black holes are
created right on the string, but that the latter does not snap or `fray'. 
This process
can be described by means of the Ernst metric with a constant conical 
deficit along the axis where the black holes lie. 
In principle the presence of the string {\it does} affect the pair creation 
rate: it is enhanced relative to the creation of black holes away from the 
string, since the action of the instanton is smaller
precisely by a factor of the conical deficit. This enhancement is no more 
than the effect (discussed in \cite{emp2} in the context of thermal 
nucleation of black holes) that a black hole
nucleates preferentially on a string, rather than on flat space.
Now, if the holes are extreme, the string cannot penetrate the horizon 
of either of the holes.  
Rather, the vortex must wrap around each of the black 
hole horizons, so that the entire configuration will look rather like two 
peas in a pod being squeezed apart. Now suppose that the created holes are
much larger than the vortex flux tube. Then the created holes will want to
pop out of the vortex. This would suggest that the rate at which two extreme 
black holes nucleate on a (non-snapping) string will be strongly 
suppressed and probably zero.
Research on this and related problems is currently underway.

\acknowledgements

The authors would like to thank Ana Ach{\'u}carro, Fay Dowker and Simon Ross 
for useful conversations, and Ruth Gregory for carefully reading the 
manuscript. A.C. was supported by NSF PHY94-07194 at ITP, Santa Barbara and 
by Pembroke College, Cambridge. J.M.A. A.-C. was supported by Wolfson College, 
University of Oxford. R.E. was partially supported by
a postdoctoral FPI fellowship (MEC-Spain) and by grant 
UPV 063.310-EB225/95. A.S. was supported by U.K.\ PPARC grant GR/L21488.

\end{document}